# Braitenberg Vehicles as Developmental Neurosimulation

Stefan Dvoretskii[1], Ziyi Gong[2], Ankit Gupta[3], Jesse Parent[4], and Bradly Alicea[4,5]

## Abstract

Connecting brain and behavior is a longstanding issue in the areas of behavioral science, artificial intelligence, and neurobiology. As is standard among models of artificial and biological neural networks, an analogue of the fully mature brain is presented as a blank slate. However, this does not consider the realities of biological development and developmental learning. Our purpose is to model the development of an artificial organism that exhibits complex behaviors. We introduce three alternate approaches to demonstrate how developmental embodied agents can be implemented. The resulting developmental BVs (dBVs) will generate behaviors ranging from stimulus responses to group behavior that resembles collective motion. We will situate this work in the domain of artificial brain networks along with broader themes such as embodied cognition, feedback, and emergence. Our perspective is exemplified by three software instantiations that demonstrate how a BV-genetic algorithm hybrid model, multisensory Hebbian learning model, and multi-agent approaches can be used to approach BV development. We introduce use cases such as optimized spatial cognition (vehicle-genetic algorithm hybrid model), hinges connecting behavioral and neural models (multisensory Hebbian learning model), and cumulative classification (multi-agent approaches). In conclusion, we consider future applications of the developmental neurosimulation approach.

## Introduction

How do we understand the emergence of a connected nervous system, particularly in terms of how it leads to neural function and behavior? One way is to infer the co-occurrence of neural cell differentiation in a model organism [1,2]. This requires a small connectome in which cell differentiation can be tracked. Even for organisms such as the nematode *Caenorhabditis elegans* [3], direct experimental observation of this process is difficult. An embodied, *in silico* system with a generalized nervous system would provide a means to both modify the developmental process and directly observe developmental plasticity. Utilizing an abstraction to study hard-to-observe questions is in fact consistent with how theoretical modeling and simulations have been used throughout the history of neuroscience [4]. We propose that Braitenberg Vehicles (BV) [5] can be used as a means to construct such simulations. Originally proposed by Valentino Braitenberg, BVs are an embodied model of a simple nervous system. The minimalist Vehicle architecture allows us to focus on the connection between an embodied connectome and its behavioral outputs. Our approach differs from Braitenberg's thought experiments in that we allow simple connectomes and associated behaviors to develop in different ways utilizing three alternate approaches that approximate different aspects of organismal development. Rather than simply assume simplistic connections between sensor and effector, we offer proof of concept for three different strategies that provide computational models of connectome development: growing connectomes and evaluating them with a genetic algorithm, using a connectionist model where multisensory associations result from behavioral feedback, to group behaviors that emerge from fixed connectomes via interactions with common stimuli. By modeling the developmental emergence of simple connectomes in an embodied context, we are able to move towards a means to developmental neurosimulation.

[1] Technical University of Munich, Munich Germany
[2] Duke University, Durham, NC USA
[3] IIT Kharagpur, Kharagpur India
[4] Orthogonal Research and Education Laboratory, Champaign-Urbana, IL USA bradly.alicea@outlook.com
[5] OpenWorm Foundation, Boston, MA USA

**Motivation**

This work is motivated by a desire to understand neurodevelopment balanced with a need to establish an *in silico* model system that allows us to simulate processes such as learning, plasticity, and the regulation of behavior. Of particular interest is a model which allows us to model global structures such as the connectome [6]. An artificial connectome that develops in the context of a controlled environment allows us to better understand various aspects of adaptive behavior. This includes both components of the networks themselves in addition to its complex behavioral outputs. Much as with biological model organisms, their digital counterparts must allow for these processes to be experimentally tractable. The BV is a good model in this regard, since it allows for a realistic amount of complexity but also provides a means to reverse engineer this complexity.

There are three benefits in choosing the Braitenberg vehicles paradigm to model neural development: a simplified structural-functional relationship, the ability to simulate an embodied nervous system, and the flexibility of modeling a heterogeneous population of agents. BVs also balance the benefits of an embodied neural system with a simplified mapping between sensors and effectors by observing the consequences of their behavior in the world. Since the mappings between environmental input, nervous system elements, and behavioral outputs are fairly explicit, we can observe the regulatory complexity of emergent behavioral phenotypes while minimizing the complexities of a biological system.

**Generalized Models of Regulation**

Our approach to developmental neurosimulation using BVs relies upon how neural networks and agent bodies emerge via different developmental stages, in addition to the emergence of collective behaviors from mature individual actions. Our software instantiations present at least two: generalized behavioral reinforcement and Hebbian learning. Behavioral reinforcement is most famously characterized through reinforcement learning techniques [7], but the core mechanism itself can be implemented using a host of other techniques [8]. For example, Hebbian learning is the dictum that “neurons that fire together, wire together” [9]. The co-occurence of particular neural units can produce spontaneous and adaptive behaviors depending on the context. Another example comes from the implementation of genetic algorithms, where fitness functions can serve to reinforce adaptive behaviors through hill-climbing on a fitness landscape [10].

More generally, our approaches involve a mechanism that allows for some form of adaptive feedback. Even in lower-capacity cognitive agents, a greater ability to model environmental conditions or interpret sensory-motor input may lend towards these agents developing into so-called good regulators [11]. This can be achieved through regulatory mechanisms for a single agent, or regulation of behaviors across multiple agents.

## Neurodevelopment and Brain Networks

The study of neural development has a long history [12]. The phenomenon of neural development proceeds from a simple group of cells to a complex and heterogeneous network of multiple functions. In vertebrates, for example, the spinal cord and brain arise from the neural tube and a subsequent process of neuronal differentiation [13]. In turn, the neural tube precursor is a sheet of undifferentiated cells. This general drive towards complexity in network topology and behavioral substrate [14] can be observed quite clearly in complex small connectomes with specialized functions such as those found in *Drosophila* mushroom bodies [15]. BV models of development allow us to implement this drive towards complexity in a digital environment where the components of the emerging nervous system can be specified and measured.

### Principles of *In silico* Connectome Development

Even in the case of an *in silico* model, it is often difficult to approximate the complexity of a connectome. Attempts to grow a connectome *in silico* using the connection rules of an adult mouse brain [16] demonstrate the difficulty of simulating a network at large scales. While scale is a major factor in this complexity, the nature of developmental (as opposed to adult) rules are often unknown. There are, however, three principles that are derived from developmental processes and constraints: an expansion of the network, an adaptive specialization of the network, and resulting structural features that reflect function. All of these principles can be expressed using BV models, and are implicit in our software instantiations.

One unique aspect of the developmental connectome is the expanding connectome of embryogenesis. This network is first established in the embryo, and results from the differentiation of pluripotent cells into neural cells such as neurons, glia, and astrocytes. The genesis of biological connectomes can be divided into two steps: the birth of neurons, and the establishment of intercellular connections. Since neurons without connections are ultimately inviable, the birth of neurons and the establishment of physiochemical connections between these cells are necessary for plasticity in a connectome [17].

### Function and Organization of Developmental Neural Networks

Another aspect of the developmental connectome is selective elimination of connections during functional refinement. In terms of developmental plasticity, the greatest degree of connectivity occurs immediately following neurogenesis. This is due to evolutionarily conserved genetic mechanisms [18]. Once exposed to the environment, these connections are pruned so that only the most active connections remain. We see this type of pruning in the visual cortex during early life-history: as neural connections are exposed to the environment, they are reinforced [19]. In an artificial context, we expect that this will result in two different types of patterned connectivity in a biologically-inspired neural network. The first are hierarchical pathways centered on a few key cells, and the second involves connectivity between disparate sets of cells from a wide variety of nervous system regions [20]. This mix of hierarchical and distributed processing allows for many of the adaptive behaviors artificial neural systems are known for.

**Patterns of Connectivity.** We can also observe hierarchical organization and disparate regional connectivity in biological organisms. Therefore, the third aspect of the developmental connectome is the structure of function resulting from developmental processes. During the process of growth and selection, a number of structural motifs emerge that are useful for robust function of the adult connectome. In *C. elegans*, the hierarchical nature of the connectome reveals a number of higher-order organization principles such as rich-club connectivity [21] and the hourglass effect [22]. These structural aspects have their origins in neural development, and in fact are the primary basis for facilitating functions such as developmental plasticity and learning. The most obvious way we can model the developmental nervous system is to use a connectionist model. Yet connectionist models also imply a wider set of physical and computational properties. According to Farmer [23], connectionist models are dynamical systems with both interactions between variables explicitly constrained to a finite set of connections and fluid connections in terms of connective strength. In applications to development, it is this latter point that becomes highly relevant.

**Development and Connectionism**

One way to understand how the developmental process shapes the brain is to model development using customized connectionist or agent-based models. In Munakata & McClelland [24], connectionist models are shown as being useful in defining developmental trajectories, critical periods, and the ontogenetic learning process [25, 26]. The last of these (ontogenetic learning) can be defined as the developmental transition between innate processes that dominate in early development and learned mechanisms that take over later in development [27]. While not all species undergo this transition at the same rate (or even at all), computational models can generate a number of potential scenarios for this type of developmental plasticity.

Perhaps more generally, development is the process by which one level of performance or competence can lead to another [28]. As the neural substrate increases in size and complexity, the organism transitions to new behavioral regimes. These are expected to be expressed as either the modifications of previous states, or new states altogether. These behavioral regimes are the product of developmental constraints, evolutionary mechanisms, and environmental challenges to the organism [29]. In at least two examples, we can see that biological intelligence is a product of dynamical systems, not just the right set of connections between neurons. For example, the work of Rinaldi & Karmiloff-Smith [30] demonstrates that intelligence can fluctuate across the ontogenetic process, and is contingent upon both the genetics of development and environmental factors. Furthermore, it is demonstrated in [31] that so-called developmental transitions in reasoning behavior can be characterized as nonlinear dynamics and represented using a computational model.

**Representational Complexity**

As the generation of complex behavior requires a more complete representation than afforded by a standard BV, we must also think in terms of representational complexity. A structural measure of representational complexity is introduced in Quartz & Sejnowski [32], and provides us with two foundational underpinnings for creating software instantiations of developmental BVs (dBVs). Development in this context is defined by a progressive increase in representational complexity and associated anatomical structures. Moreover, increases in complexity corresponds to the interactions with the structural environment. While these points are consistent with the notion of neural constructivism [33], dBVs also require some inspiration from biological innateness. Consistent with the notion that the development process is a combination of learned experience and the unfolding of innate biological processes, Zador [34] argues that neural network simulations must include innate components in order to truly exploit the computational power of biological nervous systems. We add to the conventional literature on BVs in this respect: our instantiations incorporate hybrid representations (e.g. genetic/embodied) that exceed the traditional computational substrate of neural networks.

**Model Fidelity.** In implementing dBVs, we have also attempted to bridge the gap between strong biological fidelity and models of mixed cognitive and biological fidelity [35]. This corresponds to the deep learning/swarm instantiation presented in a later section. The other two software instantiations, in addition to the general computational developmental neuroscience model, exhibit strong biological fidelity. As such they rely on bottom-up organizational principles such as a plasticity of connections and the emergence of simple behaviors. On the other hand, mixed biological-cognitive models retain a pattern of connectivity throughout their life-history trajectory (and thus a non-plastic behavioral repertoire). Yet while each agent is used to represent singular behaviors, putting them in an environment with other agents representing the same or a multitude of behaviors can result in the observation of emergent phenomena [36].

**Embodied Cognition**

Our dBV models offer an interesting opportunity to explore the origins of cognitive embodiment. As an academic discipline, embodied cognition [37-39] draws upon disciplines such as psychology, biology, cognitive science, robotics, and complex systems. Traditional views of cognition propose that the mind is not only a logical computational engine, but also operates independently of the external environment. An essential component of these computational systems are representations that can perform symbol-manipulation [40]. Embodied cognition does not eschew representations, but views them as resulting from interactions between the organism and its environment.

Particularly relevant to small connectomes that produce behavioral output, radical embodied cognition [41] explicitly rejects the role of representations, and posits that that cognition can be described solely in terms of agent-environment dynamics. These interactions can then be understood through the application of quantitative techniques such as dynamical systems analysis. In general, embodied cognition challenges the notion that the sensory world and action in that world are peripheral or auxiliary elements of cognitive processes. According to the embodied cognition view, the body and brain are interdependent in a way that enables us to approximate both the developmental process and a distributed nervous system with minimal representation [42, 43].

**dBVs and Hybrid Representations.** The use of dBVs is an attempt to reconcile a radical embodied view of representations with both innate processes (evolution and development) and higher-level representations (multisensory integration). By examining the nature of how the modalities of sensory input influence perception, behavior, interpretation, or even representation, embodied cognition expands what is seen as integral to cognition [44]. Within the general notion of embodied cognition, there are differing perspectives about the degree of representation applicable or necessary. By taking a radical embodied cognition perspective, we can investigate what elements of cognition, or perhaps proto- or pre-cognitive feedback loops, operate with minimal if any representation or symbolic manipulation taking place [41].

**Towards Action and Neuroethology.** Developmental neurosimulation via dBVs provide a unique opportunity: Braitenberg's original conception of vehicles were embodied and representation-free models of simple internal structures that result in "intelligent" behaviors. This provides an opportunity to explore the innate and plastic components of the underlying developmental neurobiology, particularly regarding the potential range of expressed behaviors available for agents with minimal cognitive representation. Additionally, dBVs may also provide greater perspective on the Neuroethology of developing individual and group behaviors, as Graziano [45] suggests that a focus on the sensorimotor loop and the study of movement behaviors more generally is key to understanding cognition as a form of intentional action.

## Platform Descriptions and Simulations

This section presents the methods used to develop the software instantiations presented in the Results section. These include descriptions of software packages, and mathematical formalisms that describe each approach to our common problem.

### BraGenBrain

The BraGenBrain approach to dBVs applies a genetic algorithm to a conventional BV. This approach allows for a population of small connectomes to be generated and evaluated in the same environment (Figure 1). These small connectomes consist of directed acyclic graphs (DAGs) that link sensor and effector. The use of genetic operators such as crossover, mutation, and selection are used to introduce developmental plasticity, while the best performing developmental trajectories are discovered using natural selection. As the BV agents move around and interact in a sandbox simulation, agents develop both implicit (nervous system) as well as explicit (behavioral) features.

**Environment and body.** The BraGenBrain environment is a *n*-dimensional box where agents exhibit attractive and repulsive movements with respect to so-called world objects. We have only conducted experiments in a two-dimensional space with one type of world objects defined as perfect circles of equal size, although more complex environments are possible. An agent body incorporates many of the classic BV elements [5]: a body core (vehicles of rectangular shape), sensors that receive signals from world objects, and motors that move the body core through the environment. The sensors and motors are connected by an artificial connectome that is generated and evaluated by a genetic algorithm. BraGenBrain uses an agent class (vehicle) supplied with a companion factory, a companion factory class can be used to construct more complex bodies. The world is filled with objects in pseudo-random positions using the Java Random generator without taking into account closeness to already existing world objects. Details of this process are shown in Table 1.

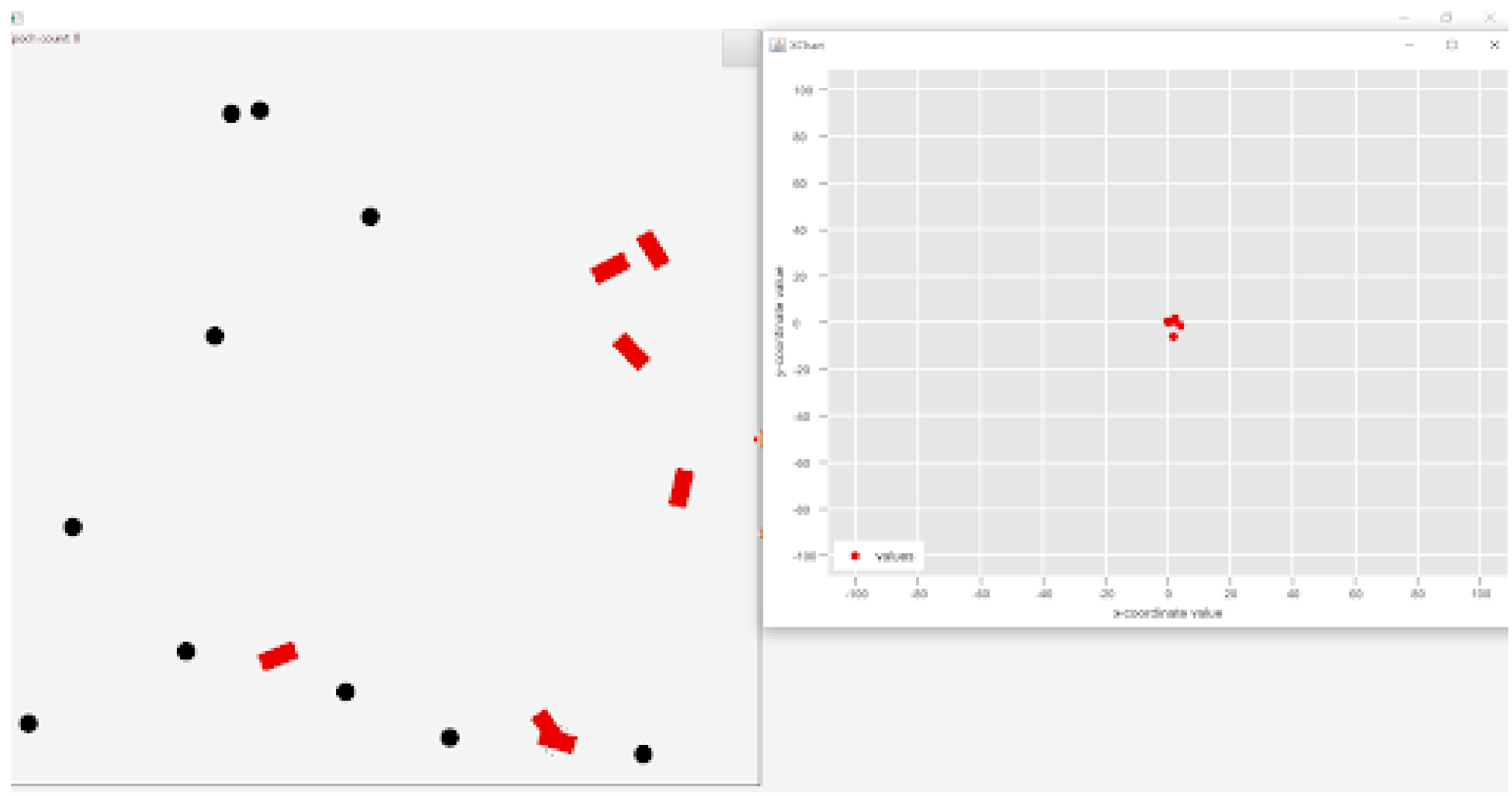

Figure 1. Screenshot of BraGenBrain environment. LEFT: spatial array with environmental features (black dots) and agents (red rectangles). RIGHT: bivariate plot of values, values for each agent are shown as red dots.

Table 1. Constructing a vehicle in the BraGenBrain environment.

```
class Vehicle {
      //internal vehicle fields and methods
      ...
      companion object Factory {
            ...
            fun simpleBVVehicle(...) : Vehicle {...}
            //other vehicle-producing functions
            ...
      }
}
```

**Vehicle movement and attraction/repulsion.** The movement of vehicles in space over time (simulation frames), coupled with their tendency to move towards and away from various stimuli, determines the fitness function upon which generated connectomes are evaluated. These components are force (*F*), rotating movement (α), and sliding movement ($tick_t$). The first component (*F*) determining vehicle behavior is to approximate the effect of objects and the environment on the vehicle's sensors. A set of environmental objects has their own effect strength taken from the random distribution between two values, which is configurable in an experiment specific manner. *F* is generated by a vehicle's forward or backward non-inertial movement, and can be calculated as

$$F = \frac{g * 100 * e}{d^2} \qquad [1]$$

where $d$ is the distance between the world object and sensor measured in pixels, $e$ is the effect strength (represented in code as *effectStrength*), and $g$ is a specific gravity constant (default value of $g = 10$).

The second step is to determine vehicle movement. Each motor produces a speed vector, which provides linear signals representing an individual motor's activity. Speed vectors can also be summed together to result in rotating and sliding movements. Rotating movement calculates the angle between two adjacent vectors using following formula

$$\alpha = arctg(\frac{y_2 - y_1}{x_2 - x_1}) \qquad [2]$$

where $x$ and $y$ are the corresponding axes coordinates of two speed vectors: the previous time point sampled ($tick_{-1}$) and the current time point sampled ($tick_0$). When the vehicle body is rotated around the body's midline, such sliding movements are calculated by adding updated *tick* speed vector lengths along the axes to the vehicle position after the previous *tick*. If the resulting angle ($\alpha$) is negative, $2\pi$ is added to rectify the value.

**Connectome Formalism.** The genetic algorithm generates new developmental variants, which in turn have a certain level of performance in a pre-defined environment. Each of these developmental variants are represented by a directed acyclic graph (DAG) with a unique topological ordering for each set of nodes (Figure 2). From each DAG, a binary string is generated from the matrix representation to represent a candidate connectome (graph) using eight bits. Each unique DAG can be defined in the form $G = (V, E)$. In this formulation, $V$ is a set of neurons and $E$ is a set of connections between neurons. Note that $\forall i, j \in V \, |E(i, j)| \leq 1$, i.e. there can not be more edges than one between a distinct pair of nodes. For the sake of balancing the signal of each neuron, we normalized the weights so each neuron output synapses weights sum up to one. We do this with a *weight function* defined as

$$w: E \rightarrow [0, 1]: \forall v \in V \sum_{u \in V} w((v, u)) = 1 \qquad [3]$$

where a function $w$ which maps from the set of edges $E$ to rational numbers between zero and one, so that the sum of weights of all edges from a particular network node is equal to *1*.

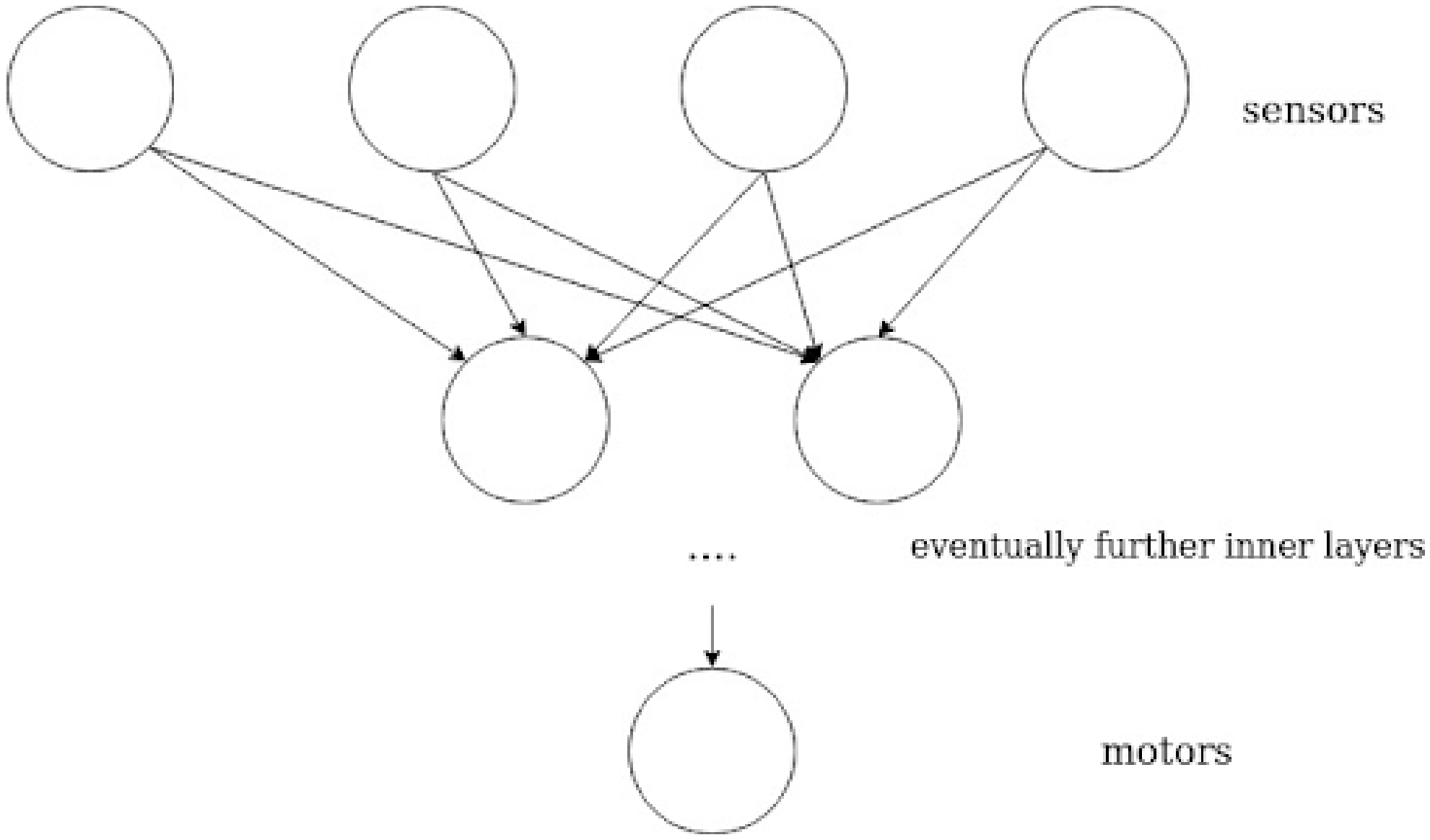

Figure 2. Schematic representation of BraGenBrain agent brain graph. Sensors and motors are connected through synapses and internal layer nodes.

There is extreme overhead in storing each connection weight in a given network as a float (4 bytes). To overcome this, we compressed the representation to eight bits, so that there were 255 possible weights. This is defined by the following equation

$$w_{concise} = \lfloor w * 255 \rfloor \quad [4]$$

in which we multiply the *w* (rational weight) of a connection by the factor of 255 and take the next lower bound integer to this number ($w_{concise}$). In this way, we preserve variety in the network while also considerably reducing the computational load. Related to this, we require a way to represent whole brain graphs in a compact manner. We chose matrix graph representation as a means to store connectivity information. This representation can be described mathematically as

$$\begin{aligned} M: m_{i,j} &= w((i,j)) \; for \; i < j \\ m_{i,j} &= 0 \; for \; i \geq j \end{aligned} \quad [5]$$

where each $i,j^{th}$ entry of matrix *M* is the weight of non-reciprocal edges between nodes *i* and *j* in our brain only if $i < j$, and zero otherwise.

**Genetic Operators.** BraGenBrain uses four genetic operators to determine various aspects of evolutionary change. The number of starting vehicles (or initial population size) can be set to a variable number, although the default value is 10. While larger initial population sizes require large amounts of computational power, small values may not yield the desired connectome diversity. The default world size is 800x800 pixels with random stimulus placement and random seed initialization. The mutation rate (default value of 0.05) introduces mutation, while the rate of random reproduction (crossover) has a default value of 0.05. The last two operators determine the fraction of vehicles selected on the basis of fitness (natural selection) or by chance (neutral drift). The default values for both forms of individual selection are 0.1 and 0.05, respectively.

**Developmental GA of the BraGenBrain Connectome**

In Figure 2, the root nodes of the connectome consist of sensors. Here, signals from the environment enter the network. The leaves of the network are represented by motors, which take inputs from the signal vectors generated by the inputs and intermediate layers. We also use a three-step evolutionary algorithm to simulate nervous system development. In the first step, mutation, random position in the binary brain representation is being mutated (i.e. *bit-flipped*) with a configurable probability. In the second step, selection keeps vehicles who exhibit the most variable movement (defined by calculating the area under their three last speed vectors) and some *lucky ones* with a certain probability. The third step is crossover, defined by randomly picking two already existing individuals, *cutting* their brain representations into two pieces in a random location and then rejoining those gametes together to produce a new brain representation. The offspring is released in the world with a new brain and a default body phenotype.

**Example of a BraGenBrain Simulation.** One set of features that make BraGenBrain particularly suitable for developmental neurosimulation is the ability to model the effects of morphological growth [46] and allometric scaling [47] in a population of behaving agents. BraGenBrain allows us to specify the size and shape of a vehicle body, the depth of a dBV neural network (the distance between sensor and effector), and the size of the neural network. Even though these are hard coded initial parameter values, they can be treated as innate features of the dBV. We can then examine how evolution unfolds on this substrate at different stages of development, or alternatively, the outcomes of different developmental trajectories.

**Evo-devo body scaling experiment.** This can be examined in two ways using an evo-devo approach. The first is to examine the effects of body shape on the effects of sensory input. To demonstrate this, we introduce three morphological scalings: a square body (1:1 scaling) and two rectangular bodies (5:4 and 3:2 scaling). In the latter two examples, the body is always longer than it is wide. The distance between sensor and effector is a square area inside the vehicle body, and can be no larger than the shortest dimension of the vehicle. This is set either as equal to the shortest dimension of the vehicle (longer distances between sensor and effector), or a value of 5 units (shorter distances between sensor and effector). This is tested for two neural network sizes: 5 and 10.

A simulation for each set of parameter values was run with identical evolutionary parameters (mutation rate = 0.05, crossover = 0.05, rate of natural selection = 0.1, rate of neutral drift = 0.05). An objective score (based on vehicle movement) was calculated for each simulation at 200, 400, 600, 800, 1000, and 1200 frames. The number of vehicles with each score at each sampling point are extracted in the form of a histogram. Each histogram provides an information-theoretic ensemble that can be reported in terms of a single $H$ value for each sampling point/experimental condition combination. The results are highlighted in Figures 3A, 3C, and 3E.

**Encephalization experiment.** The second experiment involves varying the distance between the sensor and effector with a secondary variable of neural network size. This is roughly analogous to brain size to body size scaling, which is often used as a benchmark of encephalization across species due to developmental trajectory [48-50]. This is tested for three brain size to body size relationships (1:1. 5:4, and 3:2) and two neural network sizes, 5 and 10 nodes. The same evolutionary parameters used in the first experiment were also used here. Results from the first experiment serve as the larger brain comparison here, and so H values are also used. These results are shown in Figures 3B, 3D, and 3F.

In general, the higher the H value, the more a dBV population explores its fitness landscape. Overall, there is no clear tendency for either set of experiments. Neural networks of size 10 provide more diversity in the movement-related score. Somewhat similar behaviors in the corresponding behaviors of 5- and 10-cell neural networks can be observed between shorter sensor-effector distances for the 5:4 scaling and longer sensor-effector distances for the 3:2 scaling. This type of similarity also exists for the reciprocal comparison: longer sensor-effector distances for the 5:4 scaling and shorter sensor-effector distances for the 3:2 scaling. This can be interpreted as continuities and changes in the developmental trajectory of dBVs, as the transition from immature forms to more mature forms involves both growth and asymmetric transformation of body shape.

This has a multitude of effects on behavioral exploration. One key feature is cycling between a sparse distribution of vehicle speeds to a more diverse set of vehicle speeds over time. This is likely due to the interactions between movement taxis relative to stimuli and recombination of the dBV genome. This accounts for the fluctuations over time in many of our time series. Returning to how dBVs can aid in our understanding of neurodevelopmental processes, we can look at the differences between the shorter and longer conditions for each scaling regime to understand transitions from immature (shorter) to mature (longer) phenotypes. Simulating lineages as a matter of proportional growth, we can compare Figure 3B to 3A (1:1), Figure 3D to 3C (3:2), and Figure 3F to 3E (5:4). For the 1:1 scaling, size 5 neural networks produce unstable behaviors relative to the simulation run for size 10 neural networks. This seems to be the case for both immature and mature conditions. The 3:2 scaling features size 10 neural networks that become unstable in the mature condition. Finally, the immature simulations of the 5:4 scaling feature strong behavioral fluctuations for neural networks of sizes 5 and 10. Simulations of the mature condition feature neural networks of size 10 consistently that exhibit more behavioral diversity than size 5 neural networks. For both sizes of neural network, behavior is stable across the duration of each simulation.

We conducted a two-tailed t-test for unequal variances for each pair of comparisons shown in Figure 3. Single comparisons were made for each brain size, body size combination, between neural network sizes of 5 and 10 neurons. Statistical significance was found for the following comparisons: the large condition for square ($p > 0.029$) and 5:4 scale ($p > 0.0001$) vehicles, and the small condition for 3:2 scale vehicles ($p > 0.02$). These results reveal that the effect of vehicle configuration is limited, and is perhaps transitory across different developmental configurations.

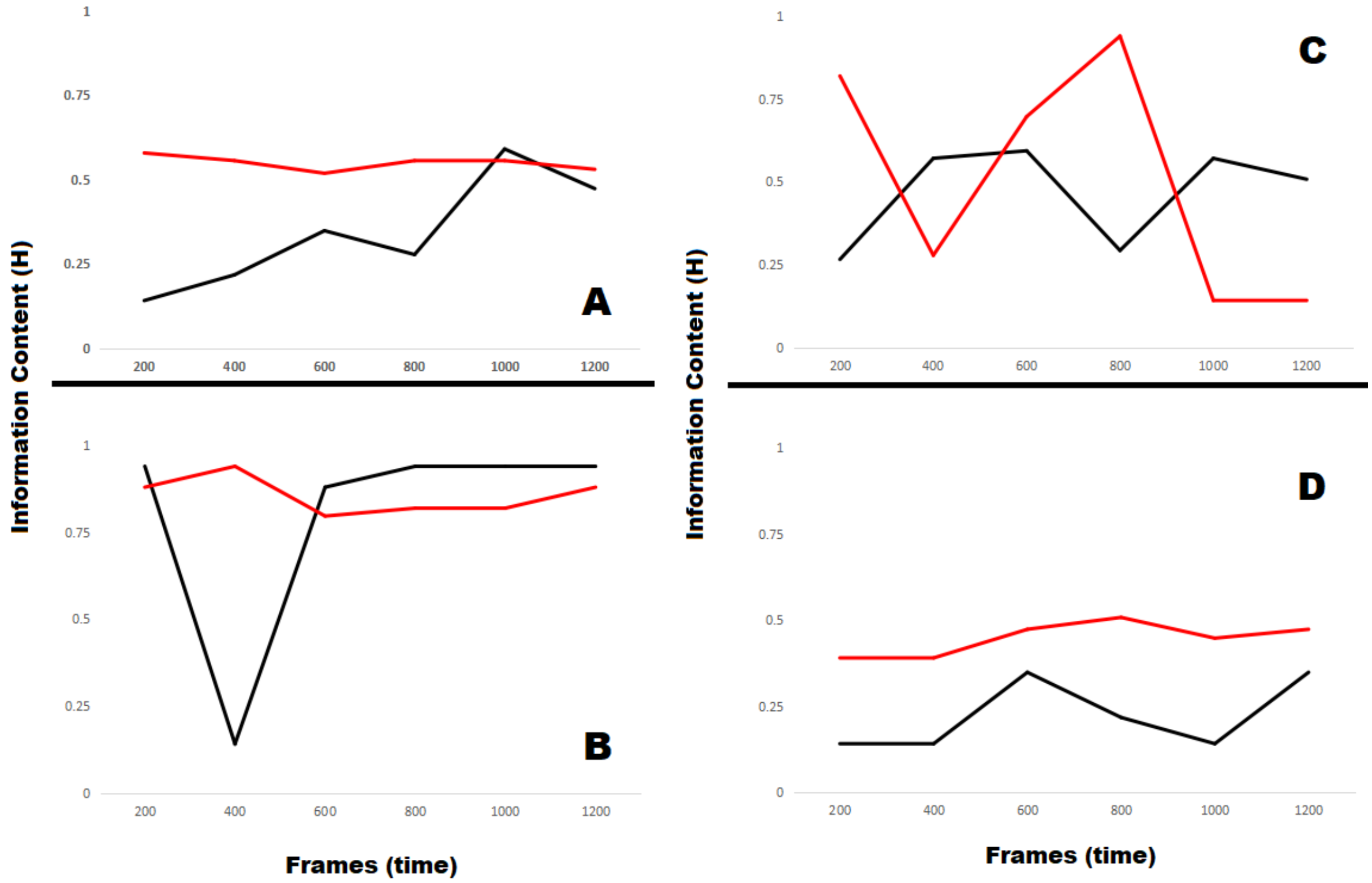

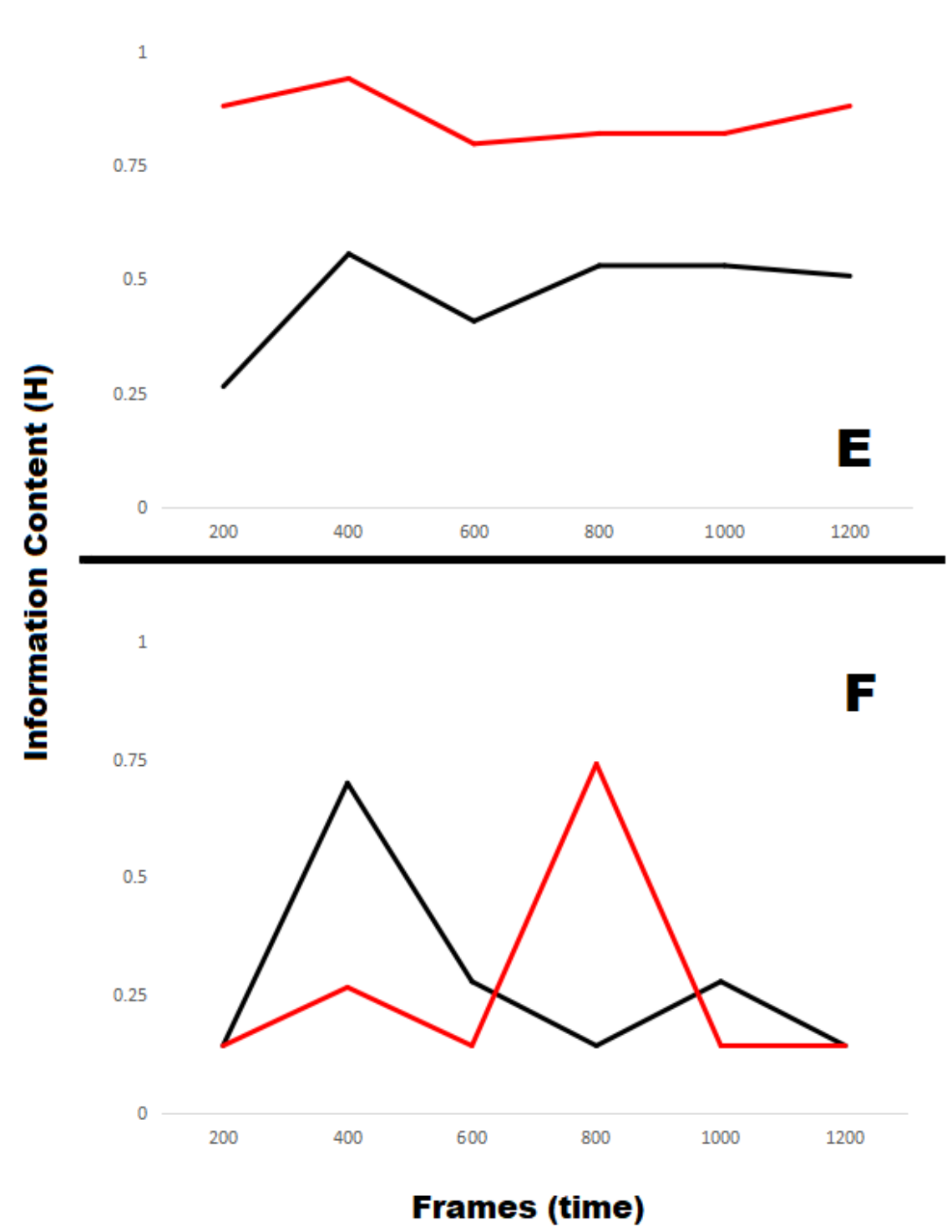

Figure 3. BraGenBrain dBVs simulated over 1200 frames for three morphological scaling values (1:1, 3:2, and 5:4) and two brain size to body size relationships (longer sensor-effector distances, and shorter sensor-effector distances). Black lines represent a neural network size of 5, red lines represent a neural network size of 10. A: 1:1 longer, B: 1:1 shorter, C: 3:2 longer, D: 3:2 shorter, E: 5:4 longer, F: 5:4 shorter.

**Open-source Code.** BraGenBrain is written in Kotlin using the JavaFX-based TornadoFX for the GUI, and is optimized enough to run agent populations up to 1000 agents smoothly on a standard personal computer. For more information, please visit the home repository: https://github.com/Orthogonal-Research-Lab/ GSoC-Braitenberg-Vehicles/tree/master/Stefan

**MultiLearn**

The MultiLearn approach to dBVs allows us to explore the developmental role of Hebbian plasticity in learning and memory. Rather than apply evolutionary principles to a population of agents, MultiLearn simulates developmental plasticity by allowing a single dBV agent to learn multisensory associations over a large number of trials. A MultiLearn dBV consists of an olfactory system (smell), a gustatory system (taste), an associative memory, a motor unit, and a judgment unit. The agent is allowed to explore its environment to discover the spatial distributions of available olfactory and gustatory stimuli (Figure 4). A single vehicle uses this information to learn various associations between taste and odor so as to approach the good sources and avoid the bad. Thus, MultiLearn dBVs are able to engage in developmental acquisition of behavioral taxis that are assumed to exist in BraGenBrain.

**Environment Setup.** An example of how MultiLearn dBVs learn is observed in the association between odor and taste in a two-dimensional space of sources that emit odor and taste. The space can be expressed mathematically as

$$O_{x,y,i} = \sum_k I^{(k)}_{o,i}\, exp\left(\frac{d^{(k)}_{d,y}}{cd_{max}}\right)$$

$$G_{x,y,i} = \sum_k I^{(k)}_{g,i}\, \theta(d' - d^{(k)}_{x,y}) \qquad [6]$$

where $O_{x,y,i}$ is the $i^{th}$ olfactory feature sensible at position ($x,y$), and $I^{(k)}_{o,i}$ is the ith feature of the odor omitted by stimulus source $k$; similarly, $G$ and $I^{(k)}_{g,i}$ are for gustatory features. $d^{(k)}_{d,y}$ is the Euclidean distance from ($x,y$) to source $k$, while $d_{max}$ is the maximum distance in space, d′ is the gustatory sensible threshold, and c is an arbitrary scalar. θ is the standard Heaviside function.

**Olfaction as a Li-Hopfield Network.** The olfactory system is implemented as a type of Li-Hopfield network [51], which is used as a standard model of olfactory bulb function (Figure 5). Li-Hopfield networks model the dynamics of two important cells in an olfactory bulb: mitral cells and granule cells. Mitral cells take in relayed sensory information from receptor cells and glomeruli as input, and produce appropriate outputs to other parts of the brain [52]. Meanwhile, granule cells serve as inhibitors of mitral cell activity [53]. In a biological context, the ratio of granule cells to mitral cells is high. In this model, however, there are equal numbers of each. The Li-Hopfield formalism can be described mathematically as

$$\frac{dx}{dt} = I + Lf_x(x) - Mf_y(y) - a_x x$$
$$\frac{dg}{dt} = I_c + Gf_x(x) - a_y y \qquad [7]$$

where $x$ and $y$ are the internal states of mitral cells and granule cells. $M$, $G$, and $L$ are the weight matrices from granule to mitral, mitral to granule, and mitral to mitral, respectively, $f$ are activation functions, $\Gamma$ is a function setting the lower triangular entries to zeros. $I$ is the input and $I_c$ is the constant ("center") input, and α is the time constant. The powers of mitral cells' oscillation are collected to be the input to the BV's associative memory.

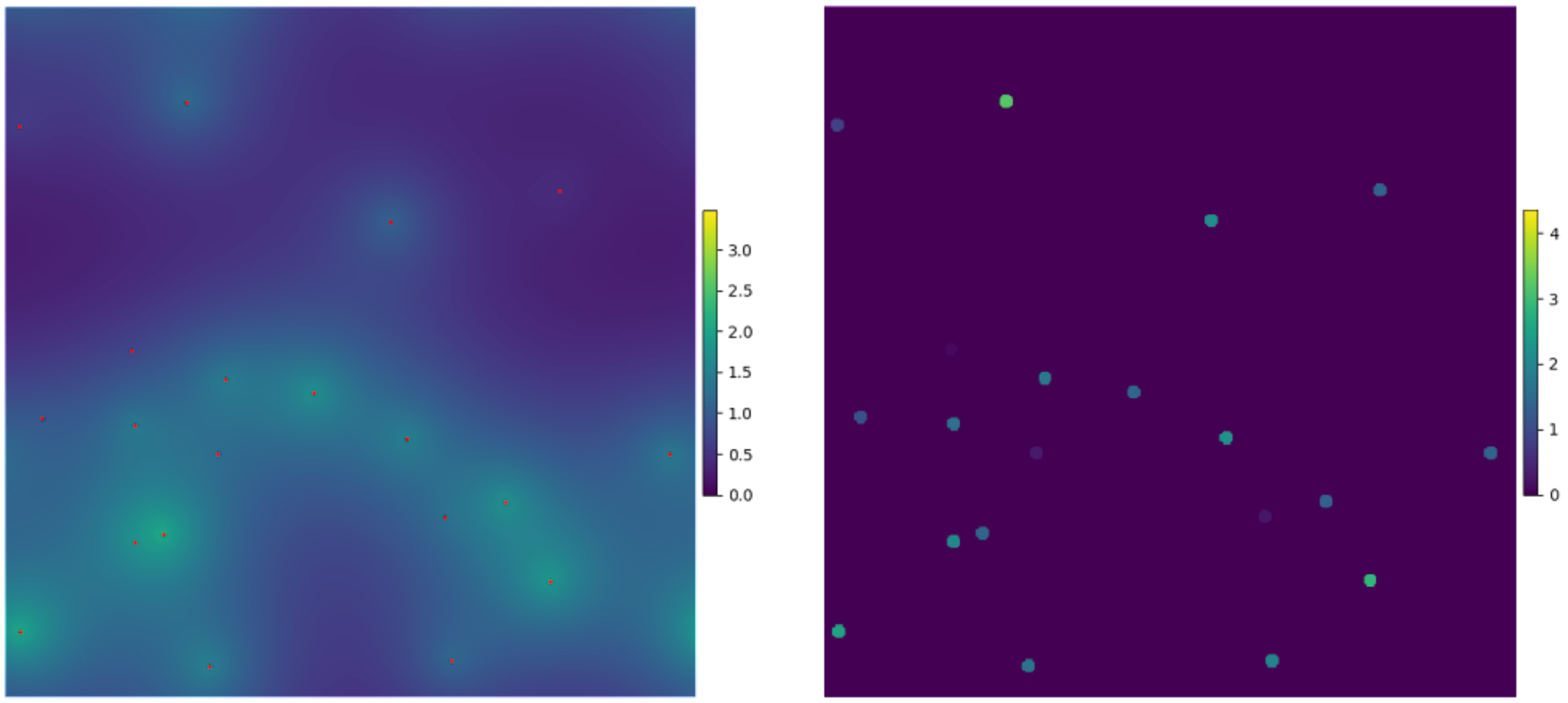

Figure 4. Odor space of one olfactory attribute (left) and taste space of one gustatory attribute (right). Number of sources = 20.

**Gustatory System.** In this model, the gustatory system is only a single layer of cells, for taste is simply an "impression" in this simulation. There is no noise involved in taste [54], or any other perturbation, so further processing of taste is redundant [55].

**Associative Memory.** To model the associative memory between odors and tastes, we implement an associative memory using the generalized Hebbian algorithm (GHA) [56] with depression. When only odor patterns are present, the associative memory maps the odor to taste (recalling mode). When odor and taste patterns coincide, the associative memory changes its weights with the following dynamics (learning mode)

$$\frac{dW}{dt} = \eta_t I'_o I_g^T - WT(I'_o I_g^T) - D$$

$$D_{ij} = \frac{\Phi}{I'_{o,j} W_{ij}} \quad [8]$$

$$\lim_{t \to \infty} \eta_t = 0$$

$$\lim_{t \to \infty} = \sum_t \eta_t = \infty$$

where $W$ is the association between $I'_o$, the processed olfactory input, and $I_g$, penalized using the depression matrix $D$ with a depression rate $\Phi$. $W_{ij}$ is set to 0 if the denominator of $D_{ij}$ is zero.

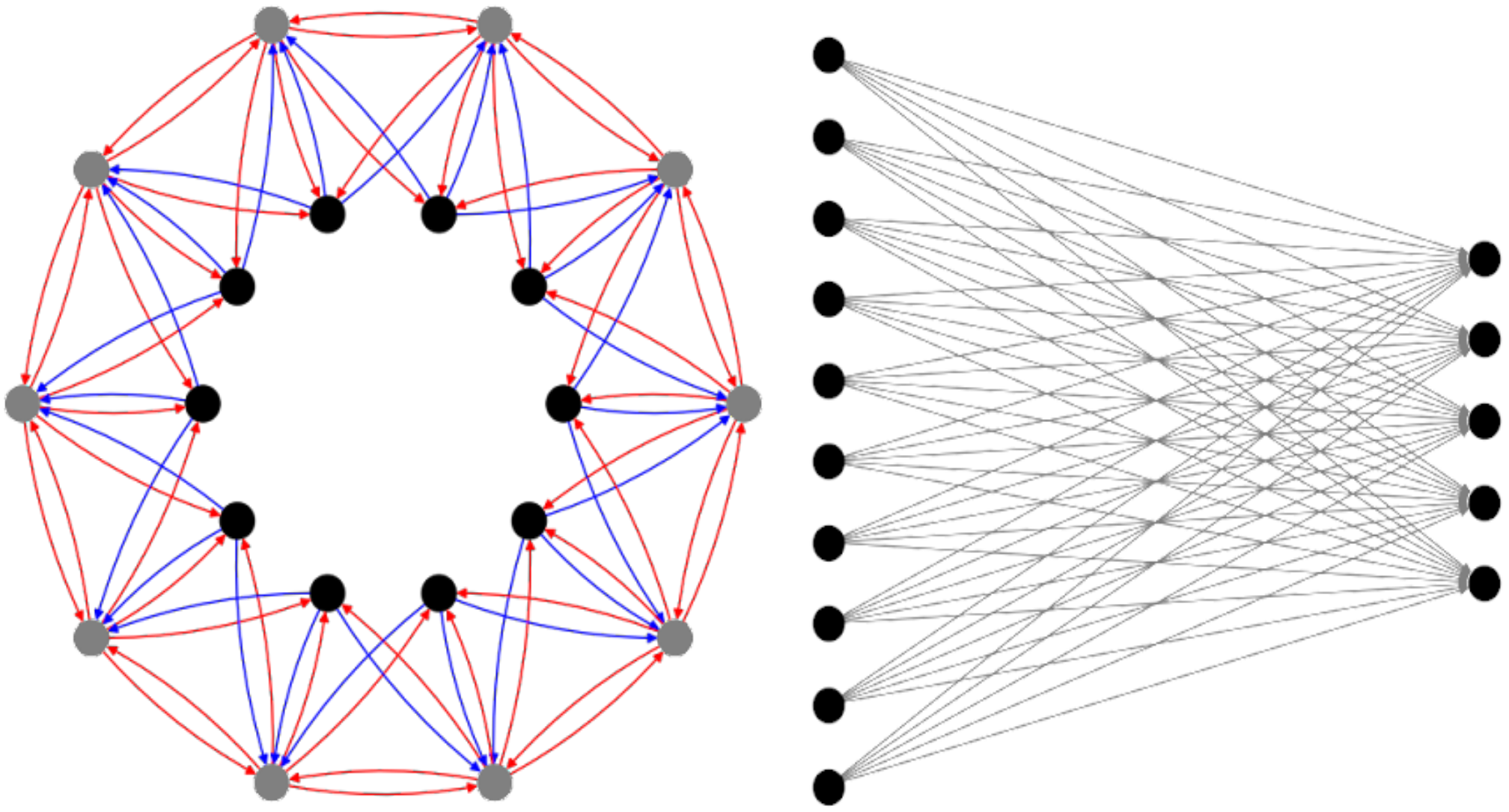

Figure 5. LEFT. Li-Hopfield model of the olfactory bulb. Mitral cells (grey) have excitatory projections (red arrow) to nearby mitral cells and granule cells (black). Granule cells simply inhibit (blue arrow) mitral cells. RIGHT. An example of an associative memory network.

The latter two expressions are required for GHA to be stable. However, this learning process, unlike typical machine learning with neural networks where samples are learned one by one, occurs in a space where samples are mixed and the time required to learn from each sample is unknown. The presentation order is randomized so that each sample can be revisited, either immediately or after the vehicle visits a sequence of samples. It is hard to determine the initial learning rate and control the pace of convergence, in addition to avoiding the effect of initial conditions when samples are not introduced serially. The learning rate is thus empirically set to a small constant $O(10^{-3})$ and decreased by 1% after the dBV gets a set of non-zero gustatory features. Moreover, the depression matrix that naively imitates activity-dependent long-term depression attempts to cancel the effect of repeatedly learning from one stimulus source and noisy data. Its effect can be demonstrated through static testing where the dBV does not move and stimuli are presented without priming.

**Motor Unit.** The motor unit is radian-based. The BV moves along the heading direction whose value is in the range: $-\pi, \pi$. When the increase in preference passes a threshold, the dBV moves forward with a little offset based on the increase. But when the decrease in preference passes the threshold, the dBV moves backward with a little offset based on the decrease. Otherwise, it moves towards a nearby source. The motor unit is implemented in *Movement.RadMotor* class. Because the learning rate of GHA has to decrease to ensure stability, the motor unit is equipped with memory to avoid repeated back-and-forth movement near the gustatory boundary of a "good" sample, which could easily lead to overfitting.

**Simple Judgment Unit.** The judgment of a source is based on its taste. A judgment unit can be defined in the following form

$$p = \sum_{i} J_i \left( I'_{g,i} \right) \qquad [9]$$

where $p$ is the preference from the summation of different judgment $J$ on recalled taste $I'_g$. If there is no recalled taste and real taste exists, then $I'_g = I_g$. The preference, the output of the judgment unit, is the sum of the output of each preference function applied to their corresponding gustatory attributes. The judgment unit is incorporated in *Simulation* class.

**Example of a MultiLearn Simulation.** To demonstrate how MultiLearn works, we construct an asymmetrical environment where olfactory stimuli decay with distances exponentially from their sources, while gustatory stimuli are sensible only when the BV is within gustatory boundaries of those stimuli. These can be represented using an odor space and a taste space (Figure 5), respectively [57]. The behaviors and association development of the dBV during its exploration of an environment are shown in Figure 6. In all simulations run ($n > 30$), the dBV successfully associates taste with smell to some degree when both taste and smell information are available. When there is no taste, it recalls the taste based on its associative memory and the smell received. In both cases where tastes were actually sensed or recalled, it can have preference on the source it is approaching, and then detour if the preferences were low. When the dBV becomes more and more mature via association, it can exhibit significant avoidance and preference behaviors, in a manner similar to small animals.

Another issue involves why MultiLearn dBVs are able to make associations and recall tastes from odors when the stimuli are mixed. Through visualization, we observed that it was able to alter its oscillatory frequencies based on changes in olfactory attributes with small latency, so that it filtered out much of the background odor and allowed the dBV to identify specifically the odor of the stimulus source the BV is approaching.

The general capacity for associating odor and taste patterns using MultiLearn dBV architectures has also been assessed. We changed the whole space (i.e. the sources and their distribution) after every $10^4$ steps, which were sufficient to let the dBV touch nearly every source at least once. After two changes, the dBV significantly lost its recallability in the new space, and manifested poor avoidance to unpreferred sources. We determined from the association matrix that it was due to the instability of our algorithm. Indeed, when the number of unique sources to learn, the order of encountering them, and the times of approaching them are all unclear, it is hard to find an optimal dynamics of the variable learning rate. Moreover, associative memories also have certain capacities that limit the amount of pattern they can encode [58].

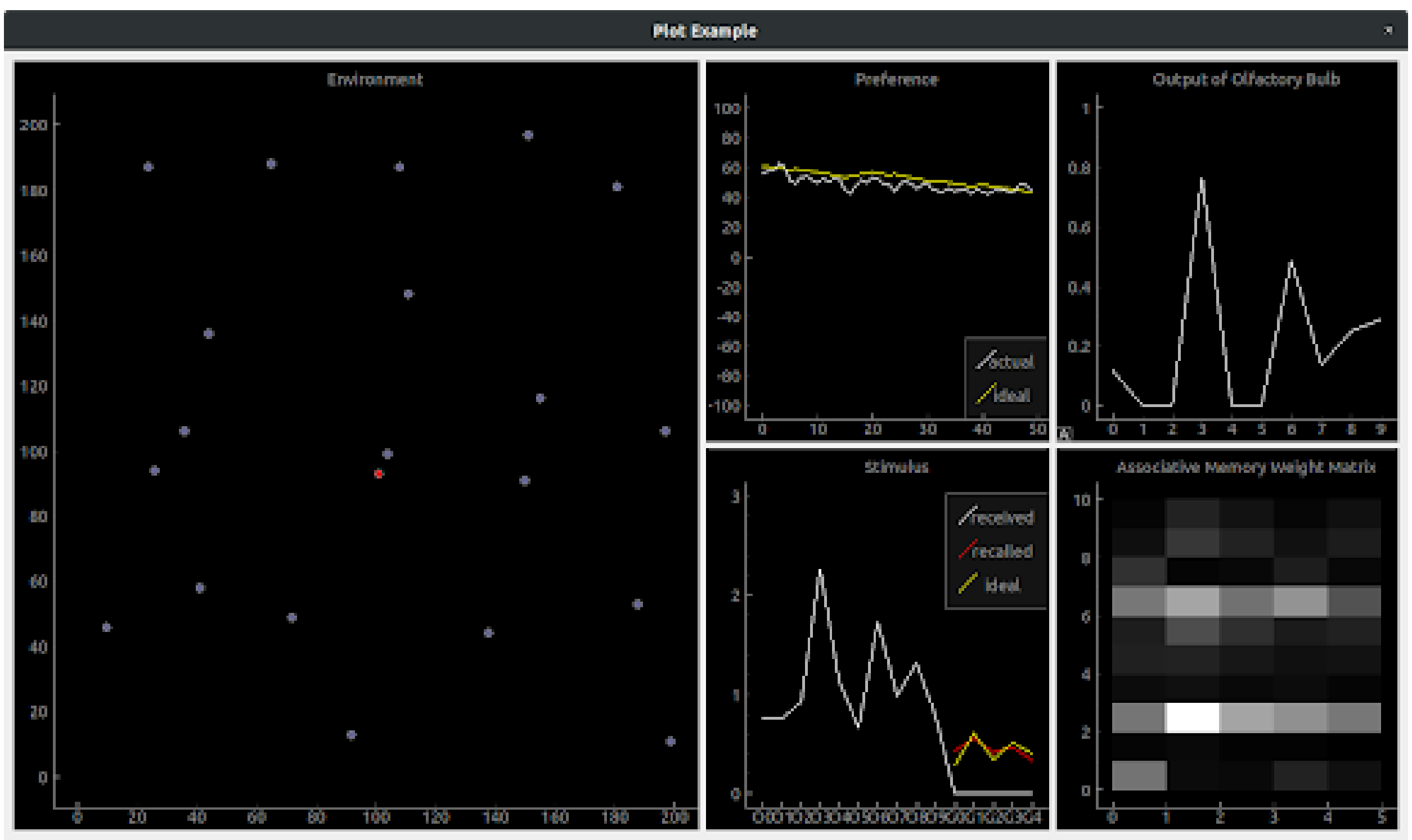

Figure 6. An example of real-time animation. Experiments are conducted using an iPython Jupyter Notebook. Left large panel: the BV (red) exploring in a space of sources (blue) emitting odor and taste patterns. Right grid, top left: the actual preference (calculated from the recalled taste pattern) compared with the ideal preference (calculated from the ideal taste pattern). Top right: the output of an olfactory bulb (Li-Hopfield model). Bottom left: the received odor pattern, and the recalled and ideal taste patterns. Bottom right: a heat map of the developing association matrix.

**Code.** This project uses Cython and C. The most time consuming parts are either written in core.c or implemented by using OpenBLAS.. Static images such as those shown above are produced through Networkx and Matplotlib, while real-time animation is generated using PyQtGraph and PyQt5. For more information, please visit the home repository: https://github.com/Orthogonal-Research-Lab/GSoC-Braitenberg-Vehicles/tree/master/Ziyi

**Methods for Simulating dBV Collectives**

Now we turn to an approach where more conventional BV agents maintain a fixed connectome, but exhibit collective behaviors when engaging with common sets of stimuli. Unlike the relatively small populations found amongst BraGenBrain agents, the dBV collectives yield emergent behaviors that result in the development of swarms of various shapes and dynamical forms. In this example, the developmental aspect refers to the emergence of collective behaviors that transcend the taxis that characterize singular BVs. In this example, each dBV is assigned a set of rules associated with environmental stimuli [5]. Based on these simple rules when multiple agents (vehicles) are introduced in the environment, they behave like intelligent swarms. There is no interaction between the agents, so their behavior is solely dependent upon the nature of the stimuli present in the environment.

**Wiring and Activation Rules.** The wiring and activation rules for the swarm intelligence approach can be defined considering functions for inhibition-dependent action

$$A_1 = \frac{k}{(k_1 + k_2 * r * r)}$$
$$A_2 = k * (k_1 + k_2 * r * r) \qquad [10]$$

where $A_1$ is excitatory activation, $A_2$ is inhibited activation, r is the distance between sensor and stimulus and $k$, $k_1$, $k_2$ are calibrated constants.

In terms of how the internal wiring acts combinatorially between the sensors and wheels, there is an identity that corresponds to a weight for each sensor-effector pairing. The weights themselves are binary (based on a threshold), and are different from probabilistic weights (based on a distribution). For lateral connections $\{w_1, w_2, w_3, w_4\}$, the weights would be {1, 1, 0, 0}. In the case of cross-talk (four contralateral connections), the weights corresponding to $\{w_1, w_2, w_3, w_4\}$ would be {1, 1, 0, 0}. This can be clearly seen in the implemented code.

**Components of Vehicle Kinematics.** As the sensory activation reaches the wheel motor, it induces two different angular velocities in the two wheels which is mainly responsible for the resulting vehicular movement. Considering the angular velocities of the two wheels and the vehicular dimensions, we can define the components of two dimensional motion of the vehicle as

$$V_x = k * \frac{\omega_1 + \omega_2}{2}$$
$$A_y = k * \frac{(\omega_2 - \omega_1) * (\omega_1 + \omega_2)^2}{\omega_1} \qquad [11]$$

where $V_x$ and $A_y$ are the components of linear velocity along the axis of the vehicle and the radial acceleration perpendicular to the axis, respectively. $\omega_1$, $\omega_2$ are the angular velocities of the left and right wheel respectively, while k is the calibrated constant depending upon vehicular dimensions. Based on an egocentric view of the environment, the vehicle can take either a left or right turn depending upon the difference in angular velocities. $\omega_1 < \omega_2$ results in a turn leftward, while $\omega_1 > \omega_2$ results in a rightward turn.

**Example of a BV Collective Simulation**

One reason to utilize Braitenberg Vehicles for developmental neurosimulation is that *synthesis* is easier than *analysis*. It is also easier to mimic behavior rather than build the true underlying mechanism that is creating that behavior. While individual Vehicle behavior is innate (hard-coded), multiple agents placed into the environment simultaneously respond to common stimuli by exhibiting a particular type of behavior. A vehicle's movement is particularly influenced by sensory activation received through its sensors and how this activation is transferred to the motor through internal wiring that mimic different types of anthropomorphic behaviors. Prolonged interaction of many Vehicles with a common stimulus leads to different emergent behaviors, often not resembling the taxis that make up the deterministic behaviors of singular agents in the collective.

In this example, some of these patterns in their static form will be shown. Figures 7 and 8 demonstrate a collective of a particular vehicle type and stimulus combination. In Figure 7, a collective of 2a vehicle types is presented with a fixed stimulus. By contrast, Figure 8 demonstrates 2b vehicle collectives presented with a moving stimulus. Each stimulus-vehicle type combination seems to produce its own set of group behaviors. While we do not present an analysis of these behaviors here, they nonetheless seem to exhibit properties consistent with emergent self-organization [59].

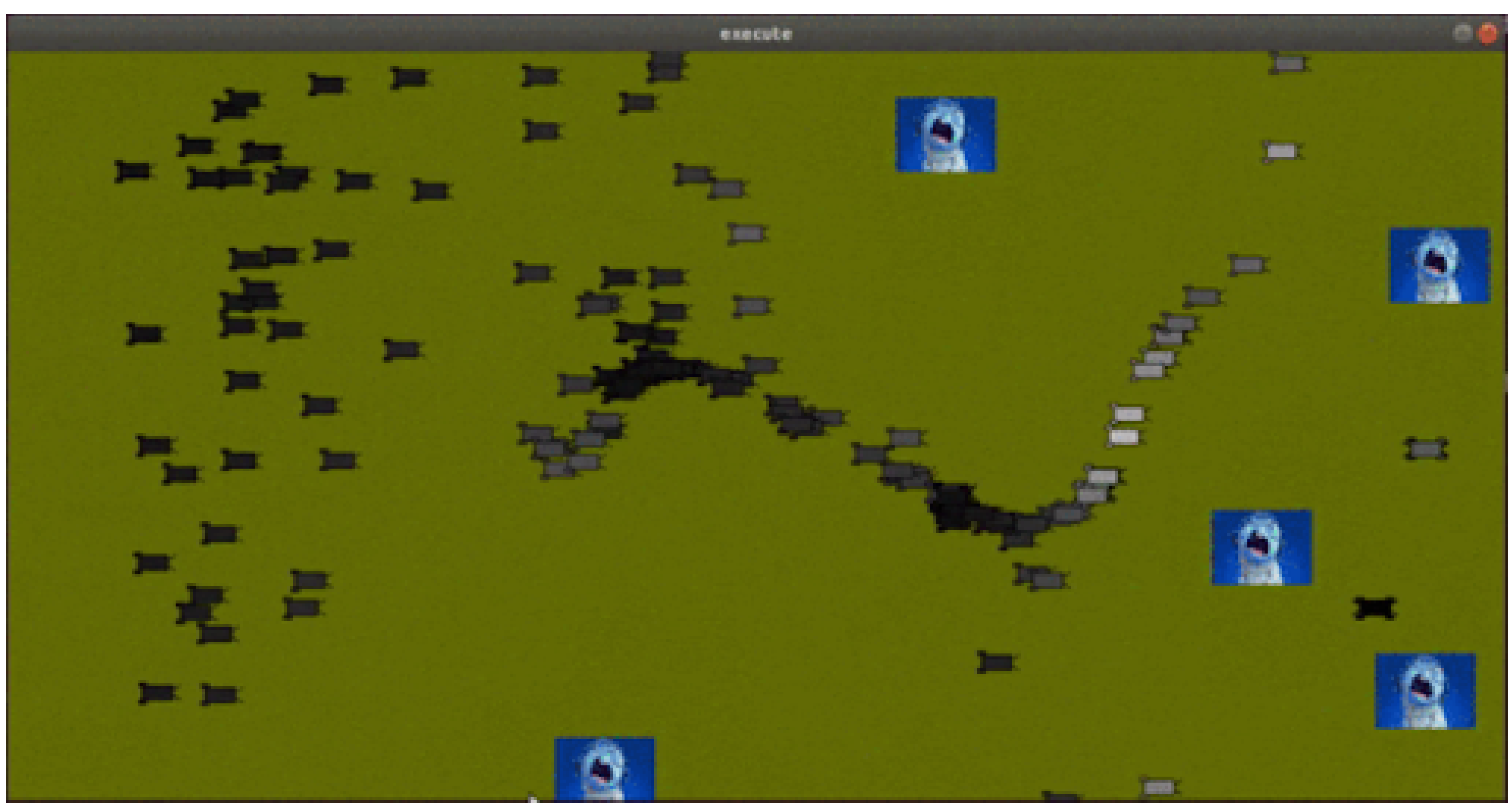

Figure 7. Screenshot of Vehicle 2a (coward) with fixed stimuli.

**Linear and Nonlinear Behavior.** Creating something that acts complex is easier than analyzing what looks like a complex system. An excitatory activation, $A_1$ is used in the case of 1A, 2A, 2B, while an inhibitory activation $A_2$ is used in the case of 1B, 3A, 3B. While monotonic functions are used for Vehicle types 1 through 3, Vehicle type 4 exhibits non-monotonic activation, as a mixture of inhibitory and excitatory connections are used. Interestingly, simulations with collectives of Vehicles types 2 and 3 can exhibit complex, nonlinear responses. Table 2 shows the relationship between vehicle type, connectivity type, and associated emotions.

**Singular and Collective Vehicle Kinematics.** In terms of vehicle kinematics for this instantiation, we find that the only thing responsible for their movement is the rotation speed of the two wheels and the difference between them. More specifically, the difference in the rotation speed of the wheels is mainly responsible for deflection from its otherwise straight-line trajectory. So, depending upon the activation received at the wheel rotators and their difference, a resultant vehicle movement is rendered. As evident from the simulation, one of the major takeaways could be that BVs are useful as the basis for configurable collectives that can develop to fit their application context. The simulations can also be used to predict/optimize paths given different kinds of neural network configuration.

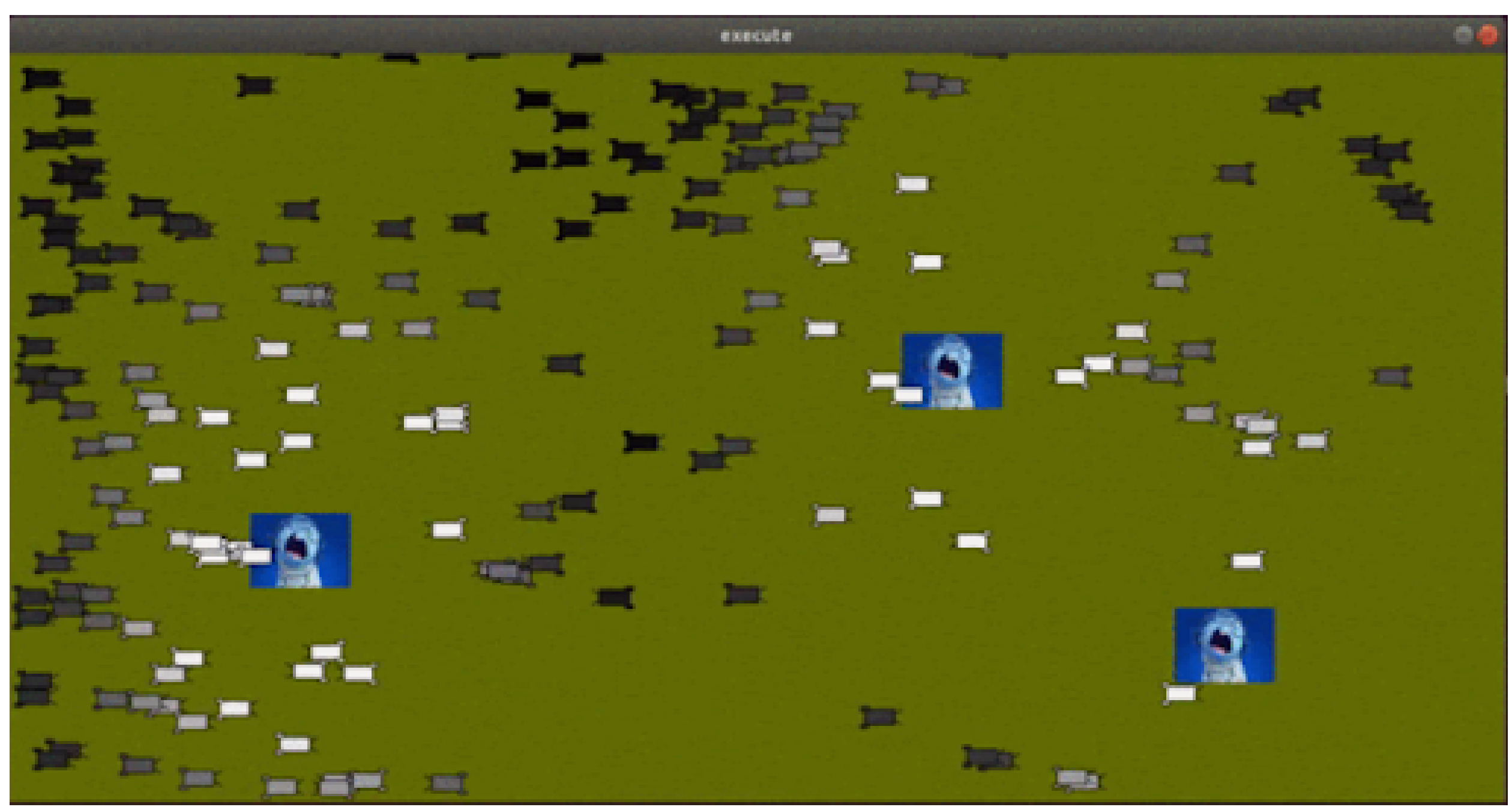

Figure 8 Screenshot of Vehicle 2b (aggressive) with moving stimuli (movement not shown).

The behavior of dBV collectives can be identified in a manner similar to what is done for patterns that emerge in the Game of Life (GoL - [60]) and for schooling fish [61]. Revisiting Table 2 suggests that there are at least two possible types of collective behavior for each emotion/connectivity type in response to a stimulus. In the fixed stimulus example, collective movements are oriented towards a single point in space, whereas for the moving stimulus collective movements may be much more complex, with twists and other complex spatial patterns.

In Figures 7 and 8, single and multiple point orientations reveal collective behavior motifs such as clusters (many Vehicles jammed together in close proximity), phalanxes (Vehicles aligned from side to side), torrents (a long curved line of Vehicles), and pinches (a short length congestion point of Vehicles). Depending on the initial density of Vehicles in the simulation, clusters can undergo a jamming phase transition [62], significantly slowing or halting Vehicle movement in response to its stimulus despite the desired behavior of individual Vehicle agents.

Table 2. Available experimental conditions to demonstrate collective BVs individual and coordinated behaviors. Vehicle type corresponds to the typology proposed by Braitenberg [5].

| Stimulus | Type | Associated Emotion | Connectivity | Observed Coordinated Behaviors |
|---|---|---|---|---|
| Fixed | 1a | Alive | Excitatory | |
| Moving | 1a | Alive | Excitatory | |
| Fixed | 1b | Alive | Inhibitory | |
| Moving | 1b | Alive | Inhibitory | |
| Fixed | 2a | Coward | Excitatory | Local jamming, Crossing-Jamming |
| Moving | 2a | Coward | Excitatory | Local jamming, Crossing-Jamming |
| Fixed | 2b | Aggressive | Excitatory | Vortex Attractor |
| Moving | 2b | Aggressive | Excitatory | Vortex Attractor |
| Fixed | 3a | Love | Inhibitory | Crossing |
| Moving | 3a | Love | Inhibitory | Crossing |
| Fixed | 4a | Instinct | Inhibitory/ Excitatory | |
| Moving | 4a | Instinct | Inhibitory/ Excitatory | |
| Fixed | 4b | Decision-making | Inhibitory/ Excitatory | |
| Moving | 4b | Decision-making | Inhibitory/ Excitatory | |

**Linking Motifs to Collective Responses.** Based on the various behaviors confirmed in Table 2, running the BV Collective simulations for different Vehicles and their corresponding stimuli yields spontaneous formation of multitude coordinated behavioral motifs. Four examples from full simulation runs are shown in Figure 9 (a-d), and are mapped to specific vehicle morphologies in Table 2. Figure 9a shows a vortex attractor, where 2b Vehicles mill around the stimulus before a few of the orbiting Vehicles quickly move towards and cross the stimulus at any one time. In Figure 9b, 3a Vehicles exhibit crossing behavior, where they make ballistic movements in all directions, crossing paths with many other Vehicles. 2a Vehicles exhibit local jamming in Figure 9c, where small dense clusters of vehicles slow down their movement towards the stimulus. Figure 9d features 2a Vehicles in a multiple stimulus world where the Vehicles exhibit a mixture of crossing and jamming behaviors. As the simulation proceeds, some defined structures transition into other structures, thus mimicking a population-wide developmental process. For example, as crossings become more dense, they may become more locally and ultimately globally jammed. Alternatively, as vehicles move away from a crossing and towards a stimulus, they may become part of a vortex attractor.

**Code.** The code for this instantiation has been constructed in the Processing.py software environment (written in Python). For the elementary setup option, opt for *python-mode*. For more information, please visit the home repository: https://github.com/ankiitgupta7/Simulations-of-Braitenberg-Vehicles

**Computational model of Developmental Neuroscience**

An ultimate goal of developmental neurosimulation is to create a generalized computational model of embodied neurodevelopmental intelligence. While this will require many more tools than currently described in this paper, such a multifaceted approach will allow us to investigate a large number of potential research questions. In general, we have found that there are three ways to approach an approximation of development and plasticity. Two of these (BraGenBrain and MultiLearn) involve a forward mapping, while the third (BV Collectives) involves an inverse mapping. The first is to use a correlation (or covariance) matrix approach, where all neural units in the nervous system are compared with every other neural unit. This results in pairwise comparisons that can lead to connectome network maps [63]. The second approach is to add nodes and arcs sequentially to a simple set of I/O connections. In this case, we get a more explicit network topology, and can observe phenomena such as preferential attachment. A third approach is to prune connections from a fully-formed network engaged in hard-wired behaviors. Using this approach, one can come to understand exactly which connections and neurons are essential for the execution of a behavior.

## Discussion and Future Plans

### Discussion

We have introduced the broader context of Braitenberg vehicles in modeling developmental processes, in addition to the application of connectionist models to understand how neural systems and behaviors are generated in development. The model instantiations presented here can be understood as large-scale models of small connectomes. Yet our three instantiations intimately tie models to behavior. To address the concerns of Eliasmith and Trujillo [64], our instantiations provide a diverse range of mechanisms that complement behaviors that emerge out of developmental processes. These mechanisms range from generalized learning and adaptation to coordinated behavior stemming from a population of fully hard-wired nervous systems. More sophisticated substrate mechanisms will be covered by use cases presented in the next section.

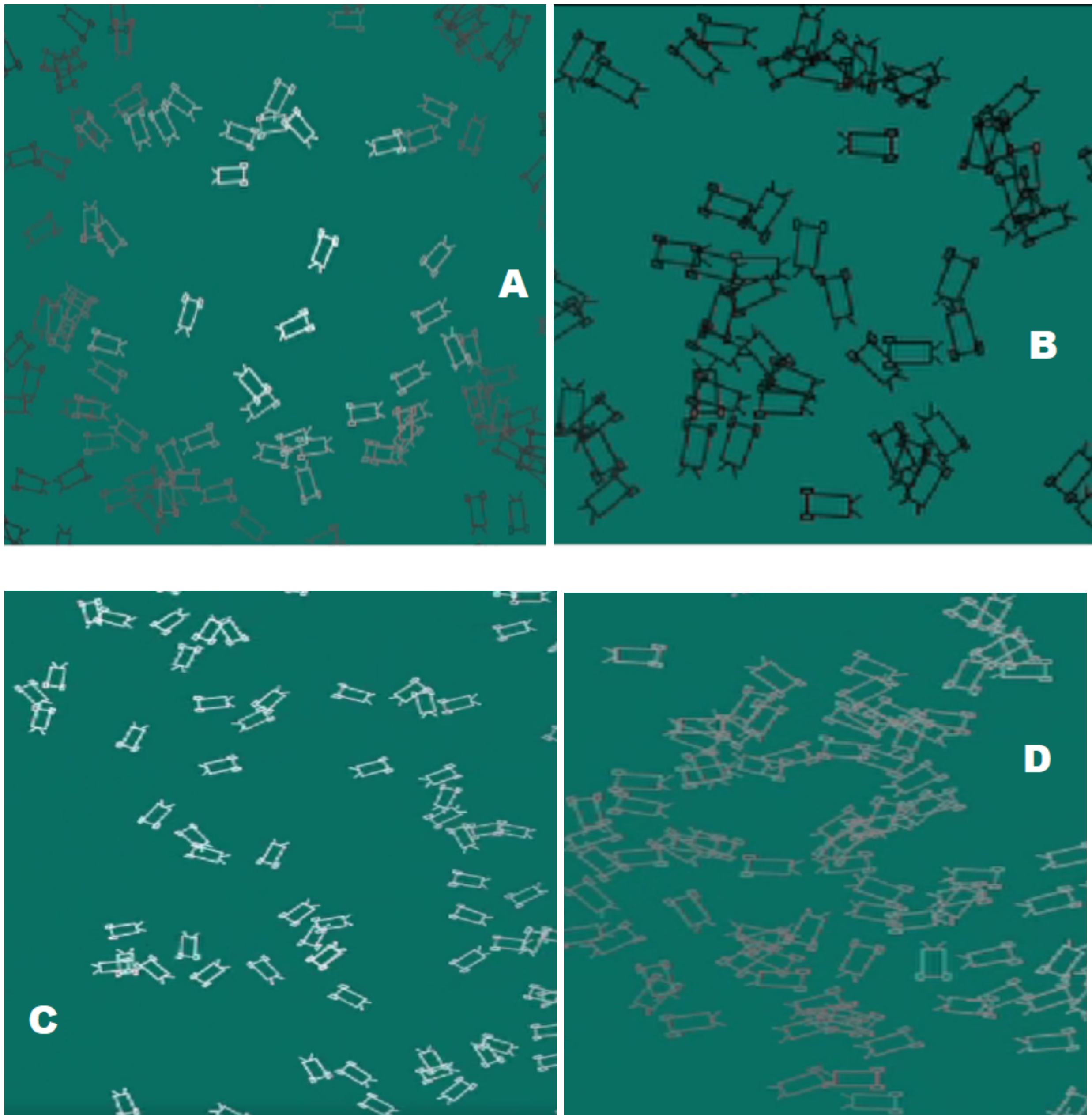

Figure 9. Characterization of motifs within Vehicle coordinated behavioral structures. A: 2b Vehicles forming a vortex attractor, B: 3a Vehicles forming a crossing structure, C: 2a stimulus Vehicles form a local jamming pattern, D: multiple 2a stimuli enable the creation of a hybrid crossing/jamming structure.

Each of the instantiations presented here are ripe for future development. As open source projects, they provide a basis for future development by the broader neural simulation community. Aside from additional functionality, these software instantiations would benefit from rigorous experimental testing. For example, experiments that demonstrate the functional range and evolvability of such models could be used for further development as well as application to basic research questions concerning brain development.

**Feedback, Regulation, and Emergence.** The brain and behavior of a single agent involves a coupling between biological structures and the environment. Developmental processes are dependent on ecological context and action [65]. Action itself (in the form of body movement) is a source of information. In fact, using action as a basis for cognition gets us around the need for a perfect internal representation, as the agent will dynamically extract information from the environment [66]. Revisiting our review of how embodied cognition is relevant to BVs, it is worth noting that agents that use minimal representation and embodied context to achieve cognitive-like behaviors also create a basis for representational structures (e.g. higher-order cognition) that in turn produce more complex behaviors [36].

We also explore emergent phenomena at two scales of biological organization: the biological connectome within an agent and coordinated behavior among a population of agents. While they result in different types of behavioral outcomes, the emergence of behaviors involves similar features that tend to be scale-invariant [67]. The first feature involves a generic form of adaptive behavior. There is a common thread between algorithmic optimization, Hebbian learning, and the mean behavior of vehicle collectives which makes all of these models exhibit interesting behavioral properties. This leads to the second feature, that of developmental contingency [68]. Each of these models relies upon sequential information that result from ever-greater amounts of interaction with the environment. Our software instantiations create developmental trajectories (either in terms of brain connectivity or group behavior) that are both unique across simulation runs and understandable through analysis. This suggests that a third and final common feature is a delineation between nature and nurture. While it is unusual to speak of either in digital simulations, we can use our software instantiations to understand the potential roles of biological innateness, environmental-induced plasticity, or the interaction of both on model phenotypes.

**Use Cases**

Building on the need to create a basis for representational structures, we can create so-called use cases to demonstrate the ways dBVs can be applied to hypothetical behavioral contexts. In this paper, a use case can be thought of as a specialized research question that matches the functionality of one of our software implementations. Here we offer three: behavioral-neural hinges, optimized spatial cognition, and cumulative classification.

**A Hinge Connecting Behavioral and Neural Models.** We have already discussed the various theoretical constructs that exist to represent neurophysiological or neuro-cognitive phenomena. Such models have been used to better understand small circuits, higher-order cognition, or phenomena such as noise and plasticity [69, 70]. On the other hand, empirical observations of animal behaviors such as predation, mating, social interaction, or puzzle solving are often model-agnostic. Therefore, we need an approach that connects models and empirical observations, and propose that models such as MultiLearn dBVs serve this purpose. We further propose that this approach acts as a hinge between biology and computation: while artificial intelligence techniques such as deep learning networks are able to generate complex behaviors [71], it is far from clear whether those behaviors are realistic. Furthermore, analytical techniques and metrics can be used to describe behaviors [72], but not create generative representations of all possible behaviors.

This relationship between models and empirical observation can be made concrete by combining our multimodal integration simulation with a meta-analysis of animal experiments focused on the mechanisms our simulation seeks to approximate. For example, in the case of neural plasticity, we might investigate multisensory inputs that include more than two senses, or perhaps multisensory integration [73] over long periods of time. Another means to implement a hinge would be to explore behaviors across the range of biological diversity using genetic algorithmic approach.

**Optimized Spatial Cognition.** We can use the BraGenBrain platform in particular as a means to study spatial cognition in unique ways. One possible approach involves using a population of BV agents to produce optimal solutions to the Traveling Salesman Problem (TSP) [74]. Our version of the TSP [75] requires an agent to find an optimal path to visit a series of locations in a region of space; in our case, a road network connecting the major cities of Europe. An agent is motivated to visit all possible paths through the presentation of stimuli embedded in a network of travel paths (roads) between nodes (cities). Each component of the potential solution is evaluated by a genetic algorithm, which subsequently results in reorganization of an agent's connectivity matrix. The agent develops a solution by reorganizing its representation of space, while this reorganization is guided along using a fitness criterion. This potentially allows for an optimal solution to TSP in a much quicker fashion than other algorithms. Moreover, it teaches us about how a spatial representation is constructed during developmental learning.

The TSP is usually characterized as a combinatorial optimization problem, where the optimal path involves visiting every location only once. Yet as a cognitive problem, this requires an agent to construct a global spatial map that can be integrated over a number of iterations. Humans can solve this problem using a number of cognitive heuristics [76], but a small-scale, developing nervous system can demonstrate how these strategies are synthesized. In the BraGenBrain implementation, we can use a fitness-based performance heuristic: paths with shorter distances can be assigned a higher fitness, which enforce efficient global solutions.

**Cumulative Classification.** While the behaviors exhibited by the BV collectives approach appear to be emergent, they do not rely upon explicit forms of coordination (communication) between agents. Instead, their behavior can be classified as cumulative. Cumulative behavior can be defined in terms of common locations that serve to homogenize behavior of many individuals. While such behaviors can be asynchronous and uncoordinated, they also result in the formation of emergent structures. We also propose a deep learning application based on an interpretation of coordinated dBV behaviors. In this application, image stimuli are presented to a population of vehicles. The vehicles are trained on this image, and display behaviors based on both their nervous system configuration and many agents responding to a single stimulus. A deep learning model is then applied to classify behaviors generated by the trained BV collectives. The stimulus is a provocation of behaviors that can be classified and further associated with the image stimuli as labels and other metadata. This will lead to associations that could inform our understanding of affective learning [77]. For example, the bark of a dog or its images to be considered as cowardness in the agents (Vehicle 2a in Table 2). This type of approach can serve as an alternative to existing approaches to classification such as multi-agent deep reinforcement learning [78].

Cumulative culture [79, 80] can also serve as a guide to explaining behaviors exhibited by BV collectives. Populations composed of the same vehicle type (Table 2) exhibit different coherent behaviors depending on whether they are presented with a fixed (Figure 7) or moving (Figure 8) stimulus. In a typical collective behavior model, there is some form of communication between the agents as they align their behavior. In this case, specific BV configurations lead to specific types of cultural traditions (see Table 2), which in turn lead to convergent behaviors. While coordination (and in turn communication) is non-existent in the BV collective instantiation, cumulative behavior nevertheless provides a means for convergent behavior. This is distinct from coordinated collective behavior such as that observed in bird flocks, fish schools, and insect swarms [81-83]. We can enforce intra-agent interaction by implementing interaction rules [84-86] that promote intra-agent interaction. In any case, BV collectives provide an opportunity to observe and analyze emergent behavioral modes by presenting different stimuli types to provoke behavioral variety.

**Limitations and Future Issues**

One issue we face is the limited biological realism of our models. We use abstractions of concepts such as plasticity, innateness, and embodiment to characterize this process. Yet do our models really capture something that would be useful to understanding the development of nervous systems? Our instantiations resemble so-called toy models of evolutionary processes [87], which allow us to represent biological processes in a simplistic manner.

Another issue that we face in building developing networks (where nodes are added dynamically) is the issue of modularity. In complex brains across a wide range of organisms, mature brains tend to be modular, or divided into structural and functional subcomponents. This is both consistent and inconsistent with what we observe in our developmental models. In the case of the BraGenBrain implementation, partial solutions do not necessarily equal the global solution. This means traditional approaches to emergent complexity in genetic algorithms such as the building block method [88] are not suitable for building progressively larger brains.

In conclusion, this work leads in a number of promising future directions. These might include scaling up the complexity of simulated nervous systems, or adding higher-order representational systems such as semantic kernels [89]. In the case of the former, we found that even moderately large numbers of neurons are computationally implausible using conventional forms of representation. This limitation might be overcome with a combination of high-performance computing techniques and multicore algorithms. We can also overcome the limitations of size by making each neuron richer in terms of the information it conveys. Adding a semantic component to small connectomes might provide an alternative means of understanding how mental concepts emerge from brains.